\documentclass[prd,aps,twocolumn,showpacs]{revtex4}
\usepackage{epsfig}
\usepackage{bm}
\def\pd#1#2{{\partial #1\over\partial #2}}

\begin{document}

\title{Decoupling and coherent plasma oscillations around last scattering}

\author{\small  A. Bershadskii}
\affiliation{\small {\it ICAR, P.O.\ Box 31155, Jerusalem 91000, Israel}}

\begin{abstract}
Coherent properties of the baryon-photon fluid decoupling are considered 
in the terms of an effective nonlinear Schr\"{o}dinger equation for a macroscopic wave function 
that specifies the index of the coherent state. 
Generation of a transitional acoustic turbulence preceding formation of large-scale condensate in the plasma 
and its influence on the CMB power spectrum has been studied. A scaling $k^{-1}$ law is derived for the 
CMB Doppler spectrum $E(k)$ (angle-averaged) in the {\it wavenumber} space, for sufficiently large wavenumber 
$k$ and for the weak nonlinear and completely disordered initial conditions. Using the recent WMAP data 
it is shown that the so-called first acoustic peak represents (in a compensated spectral form) 
a pre-condensate fraction of the spectrum $E(k)$ at a rather advance stage of the condensate formation process.

\end{abstract}


\maketitle

\section{Introduction}
The recent high resolution cosmic microwave background (CMB) radiation measurements 
(see, for instance \cite{wmap}) 
provide a possibility for quantitative investigation of
the nontrivial processes taking place
at the {\it decoupling} of the baryon-photon plasma (fluid) in the early universe. Before the decoupling the 
CMB photons were tightly coupled to the baryons by photon scattering (mainly from the 
free electrons) forming so-called baryon-photon fluid. At a certain stage of the early universe 
expansion the energy of the CMB photons becomes insufficient to keep hydrogen ionized and a recombination
process becomes dominating in the baryon-photon fluid. The recombination is a gradual process, 
but it serves as a trigger and a catalyst of the decoupling of the baryon-photon fluid. Before the 
decoupling the baryons and photons were tightly coupled whereas after the decoupling the photons are 
effectively free. Physical processes at the decoupling itself can be described on different levels. 
In present paper we will concentrate on the {\it coherent} properties of the decoupling.    
It is well known that the main energy of the plasma fluctuations at a certain (advance) 
stage of the decoupling process is concentrated in the coherent 
states (the acoustic coherent oscillations - stochastic standing waves). Originally the 
acoustic oscillations become phase-coherent when the causal horizon overtakes their wavelength. 
Around the last scattering surface the radiation density and 
ionization are high enough that the photon drag on the baryons can be still significant. Therefore, 
the baryons are involved in the coherent motion by the photon drag, which differentially 
accelerates the ions and electrons. The resulting significant electron-ion drift generates large-scale 
{\it coherent} electric currents and fields, which rather quickly become comparable to the photon drag 
in their effect on the electrons (see, for instance, \cite{bd},\cite{ho}). While we can readily introduce 
the photon (Thomson) drag in the dynamical plasma equations it is rather nontrivial task to describe the 
coherent (collective) phenomena at the decoupling, which are crucial for the entire process. To tackle this problem we 
suggest to use an effective nonlinear Schr\"{o}dinger equation for macroscopic wave function, which is a 
complex-valued classical field that specifies the index of the coherent state. To the leading order each 
coherent state evolves along its 'classical' trajectory which is given by the corresponding nonlinear 
Schr\"{o}dinger equation. The idea is based on the Ehrenfest's theorem stating that the the expectation 
values of displacement and momentum given by the Schr\"{o}dinger equation obey time evolution equations 
which are analogous to 'classical' ones in the the Madelung representation. In this representation 
the macroscopic wave function of the free (non recombined) electrons can be defined as 
$\psi = \sqrt{n_e} \exp[{\rm i}S({\bf x},t)/h_c]$, 
where $n_e$ is the electron number density, $\nabla S({\bf x},t)=m {\bf v} ({\bf x},t)$ 
with $m$ and ${\bf v} ({\bf x},t)$ as the mass and velocity of the electrons, $h_c$ here is not the 
Plank constant but a parameter characterizing coherence in the system (real quantum effects are 
negligible in the considered situation, cf recent papers \cite{gri},\cite{hab} and references therein). 
Then, momentum equation for the electron can be considered as a nonlinear Schr\"{o}dinger 
equation for the macroscopic wave function $\psi ({\bf x},t)$:
$$
{\rm i}h_c \pd \psi t = -\left[ {\frac{h_c^2}{2m}} \nabla^2  +\eta ({\bf x}) 
- U_{eff}(|\psi|^2) \right] \psi,   \eqno{(1)} 
$$
where $\eta ({\bf x})$ is a real-valued potential, $U_{eff}(|\psi|^2)$ is an effective potential 
represented by a real-valued function of the free electrons density $n_e=|\psi|^2$. Analogous approach 
at rather different circumstances was used for a mean field description of dense Fermi plasmas 
(see, for instance, Refs. \cite{m},\cite{se},\cite{ss}). Unlike the quantum plasma considered in Refs. \cite{m},\cite{se},\cite{ss} the original (input) coherence at the decoupling comes from an external source 
(cf Section VI). 
At this description the addition of the effective potential $U_{eff}(|\psi|^2)$ to the equation (1) 
is possible due to a specific gauge invariance of the equation. 
Namely, it can be readily shown that addition of the potential $U_{eff}(|\psi|^2)$ (satisfying condition 
$\lim_{n \rightarrow 0} \int U_{eff} (n) dn=0$) does not affect the 'classic' limit (i.e. the Ehrenfest's theorem) 
$$
m d\langle {\bf r}\rangle/dt= \langle {\bf p} \rangle,~~~d\langle {\bf p} \rangle/ dt = - \langle \nabla \eta \rangle  \eqno{(2)}
$$
where ${\bf p}= m {\bf v}$ is momentum. The effective potential can be also directly depending on ${\bf x}$, 
but the potential $U_{eff}({\bf x},|\psi|^2)$ should be slowly varying on ${\bf x}$ (in the adiabatic 
terms of the Ehrenfest's theorem). About a possibility of nonlocal dependence of $U_{eff}(|\psi|^2)$ on $|\psi|^2$ 
see Section VI. It will be shown below that such type of invariance of the Schr\"{o}dinger equation 
is crucial for the condensation process.  We deliberately have the dissipation term not included
into the equation because we will concentrate on the
large- and intermediate-scale processes, while the dissipation
term is usually related to the small scales (we will return to this problem below). 
The first term in the right-hand side of Eq. (1) provides, in particular, a possibility of tunneling at the decoupling. 
On the other hand, it is believed that the nonlinear Schr\"{o}dinger equation can describe development of 
the coherent regimes \cite{ks}-\cite{bs} at a certain stage of evolution. 
Theoretical considerations (see for instance, \cite{ks},\cite{kss}) show that this regime 
can appear in a long-wavelength region of wavenumber space after the breakdown of the regime 
of weak turbulence: the Kagan-Svistunov (KS-) scenario. 
This is supported by recent numerical simulations for the weakly interacting Bose gas \cite{bs} and for the dense 
Fermi plasmas \cite{ss}. 
It is expected that there exist three different stages in the time development for the KS-scenario: 
weak turbulence, transitional turbulence and condensate.  A pre-condensate begins its formation in a 
long-wavelength region at the transitional turbulent regime. This transitional turbulent regime 
is the most difficult for theoretical description \cite{bs}. 
There is a problem to define a continuous vorticity filed in the quantum-like turbulence due 
to non-rotational nature of the velocity field defined on the wave function given by the Schr\"{o}dinger equation. 
Velocity field defined on the wave functions is a potential field. Vorticity of such field 
vanishes everywhere in a single-connected region. It is believed that in the condensate state itself 
all rotational flow is carried by quantized vortices (the circulation of the velocity around the 
core of such vortex is quantized). This idea turned out to be very fruitful, and recent experiments 
give direct support for existence of such vortices in the condensate \cite{Donnelly},\cite{vn},\cite{sreeni}. 
However, for the transitional turbulence apparent absence of a continuous field associated with vorticity 
is a difficult problem. Moreover, it is clear that the only flux of the particles from the region of lager energies
toward the future condensate in not sufficient for transformation of the completely disordered 
initial weak turbulence into the condensate with its local superfluid order and 
with the tangles of quantized vortices. A certain vorticity-like quantity have to be involved in 
this process just before appearance of the tangles of quantized vortices.  
In present paper we study a generalized vorticity defined on the {\it weighted} velocity field. The 
generalized vorticity is not vanishing in the bulk of the flow. Then, we show that adiabatic invariance 
of an enstrophy (mean squared generalized vorticity) 
controls the transitional turbulence just before formation of the condensate with its 
tangles of quantized vortices. 
Since in the above described scenario the transitional
turbulence already corresponds to a coherent
regime (in this regime the phases of the complex amplitudes
of the field $\psi$ become strongly correlated) we will speculate that an advance stage of the decoupling can
be associated with the transitional turbulence regime of
nonlinear Schr\"{o}dinger equation (1) . Them we will consider
significance of the generalized vorticity for the CMB
photons visibility function, which is the main CMB characteristic
of the decoupling stage. This scenario provokes
an additional speculation: to associate
the condensate regime itself (with its tangles of quantized
vortices - the topological defects) with the universe state just after the decoupling. 
The recombination process considerably reduces the number of the free baryons 
before the last stage of the decoupling but a sufficient number of the 
free baryons can still exist at this stage in order to form a condensate due to the 
{\it fast} character of the condensation process (Section VII and Fig. 5).  
The far reaching consequences of such additional speculation
are quiet obvious, cf for instance \cite{gam},\cite{pietr},\cite{bp}. 

\section{ Generalized vorticity}

Let us recall certain well known properties of velocity field defined on the wave function. 
In order to calculate the expectation value of velocity one can evaluate the time 
derivative of space displacement $\langle {\bf r} \rangle$:  
$$
\langle {\bf v}\rangle = \frac{d \langle {\bf r} \rangle}{dt}= \frac{d}{dt} \int {\bf r} |\psi|^2 d{\bf x}= 
 \int {\bf r} \pd {|\psi|^2} t d{\bf x}  \eqno{(3)}
$$
Substituting $\pd {|\psi|^2} t$ provided by the Eq. (1) one obtains
$$
\langle {\bf v}\rangle = \int \frac{{\rm i}\hbar}{2m} [\psi \nabla \psi^* - \psi^* \nabla \psi] d{\bf x} \eqno{(4)}
$$
where we have integrated by parts with corresponding vanishing or periodic boundary conditions. The integrand in the right-hand-side of Eq. (4) is the {\it real}-valued distribution of the quantum-like velocity 
$$
{\bf j} = \frac{{\rm i}\hbar}{2m} [\psi \nabla \psi^* - \psi^* \nabla \psi]  \eqno{(5)}
$$

The real-valued quantum-like velocity filed itself one can obtain comparing the definition $
\langle {\bf v}\rangle = \int {\bf v} |\psi|^2 d{\bf x}$  
with Eq. (4), that results in
$$
{\bf v} = \frac{{\rm i}\hbar}{2m}~ \frac{[\psi \nabla \psi^* - \psi^* \nabla \psi]]}{|\psi|^2}= 
-\frac{{\rm i}\hbar}{2m}~\nabla \left[ \ln \frac{\psi}{\psi^*} \right]   \eqno{(6)} 
$$
i.e. the quantum-like velocity (6) is a potential field and $curl~{\bf v} = 0$ everywhere in a single-connected 
region.\\

Let us now define generalized vorticity on the weighted velocity field as 
$$
\omega = \frac{curl~ (|\psi|^2 {\bf v})}{|\psi|^2} = curl~{\bf v}+ \frac{\nabla |\psi|^2 \times {\bf v}}{|\psi|^2}  \eqno{(7)}
$$
First term in the right-hand-side of Eq. (7) is the {\it ordinary} vorticity. 
When there is no the quantized vortices this term is zero due to Eq. (6). In this situation the second term 
in the right-had-side of Eq. (7) determines the generalized vorticity. In this case ${\bf \omega}$ is {\it orthogonal} to the "plane of motion", which is stretched over the vectors: ${\bf v}$ and $ \nabla |\psi|^2$, at any point of the space. In this sense the transitional turbulence can be considered as a locally two-dimensional one.

The second ({\it linear}) term $\eta ({\bf x})\psi$ in the right-hand-side of Eq. (1) does not affect our further results and will be omitted for simplicity. Dynamical equation for the generalized vorticity similarly to the dynamic equation for the density $|\psi|^2$ (in a dimensionless form, for simplicity)
$$
\pd {|\psi^2|} t = {\rm i} [ \psi^* \Delta \psi - \psi \Delta \psi^*] \eqno{(8)}
$$
does not contain the {\it nonlinear} terms explicitly
$$
\pd {~{\bf \omega} |\psi^2|} t = [\nabla \psi \times \nabla (\Delta \psi^*)] + 
[\nabla \psi^* \times \nabla (\Delta \psi)]  \eqno{(9)}
$$(the right-hand-side terms in the Eqs. (8),(9) come from the linear term of the Eq. (1) only). 
This cancellation of the nonlinear terms in the dynamical equation for the density, Eq. (8), 
results (after integration by parts) in the conservation law 
for the total number of the particles: $N = \int |\psi|^2 d{\bf x} =const$. The cancellation of the nonlinear terms in 
the dynamical equation for vorticity, Eq. (9), results in a less strong result. Namely, the enstrophy (mean squared generalized vorticity) turned out to be an adiabatic invariant for the transitional turbulence.

\section{Adiabatic invariance and scaling transitional regime}
 
According to the theoretical predictions \cite{ks},\cite{kss} if one starts from a self-similar 
solution of the equation (1) for weak-nonlinear conditions, then the so-called coherent regime \cite{ks},\cite{kss},\cite{ly}, will be developed at a certain stage of evolution. The first stage 
of evolution leads to an explosive increase of occupation
numbers in the long-wavelength region of wavenumber space where the ordering process takes place. 
From the beginning of the coherent stage of the evolution we will call the long-wavelength region 
of wavenumber space as pre-condensate fraction, while the rest of the wavenumber space we will 
call as above-condensate fraction. 

At a certain time, close to the blow-up time of the self-similar solution, 
the coherent regime sets in. After this time the system has a certain transitional turbulent period. 
At the end of this transitional period appearance of a well-defined tangle of 
quantized vortices indicates the final (condensate) stage of the evolution. Therefore, 
generalized vorticity exchange between the pre-condensate and above-condensate fractions in the transitional period 
seems to be crucial for the process of the condensate formation in the KS-scenario. 

In the case of space isotropy it is convenient to deal with an 
angle-averaged occupation numbers spectrum $N_k$ in variable $k = |{\bf k}|$ in the wavenumber space
$$
N =\int N_k dk  \eqno{(10)}
$$
The total number of particles associated with the pre-condensate fraction
is $N^{pc} = \int_{k'< k_c} N_k' dk' $, where $k_c$ is the wavenumber scale separating (approximately)
the pre- and above-condensate fractions (for its determination from numerical simulations see next section). 
Then the total number of particles associated with the above-condensate fraction is: $ N^{ac}=N-N^{pc}$.
From the dynamical conservation law for the total number of particles, $N = const$, we obtain
$$
\frac{dN^{pc}}{dt} =- \frac{dN^{ac}}{dt}  \eqno{(11)}
$$
Then we can define exchange rate of the particles number, $\varepsilon_n$ 
and as
$$
\varepsilon_n = \left| \frac{dN^{ac}}{dt} \right|=\left| \frac{dN^{pc}}{dt} \right|  \eqno{(12)}
$$

Let us also consider angle-averaged enstrophy spectrum $\Pi_k$
$$
\langle \omega^2 \rangle = \int \Pi_k dk  \eqno{(13)}
$$
Then, the enstrophy associated with the pre-condensate fraction
is $\Omega^{pc} = \int_{k'< k_c} \Pi_k' dk' $, whereas enstrophy associated with the 
above-condensate fraction is: $ \Omega^{ac}=\langle \omega^2 \rangle -\Omega^{pc}$. 
The intensive particle flux to the long-wavelength region 
of wavenumber space (where the pre-condensate is formed) involves the enstrophy one. 
Unlike the total number of the particles, the average enstrophy $\langle \omega^2 \rangle$ is not 
an exact invariant of the motion. Though, characteristic time scale of the enstrophy exchange between 
the pre- and above- condensate fractions is expected to be much smaller than the characteristic 
time scale of the $\langle \omega^2 \rangle$ evolution  \cite{ks}-\cite{s}. 
Therefore, the $\langle \omega^2 \rangle$ can still 
be considered (approximately) as an 'adiabatic integral' for the exchange process. As it is 
used for the adiabatic processes this statement can be formalized as following:
$$
\frac{\Omega^{ac}}{|d\Omega^{ac}/dt|} \ll \frac{\langle \omega^2 \rangle}{d|\langle \omega^2 \rangle/dt|},
~~~ \frac{\Omega^{pc}}{|d\Omega^{pc}/dt|} \ll \frac{\langle \omega^2 \rangle}{|d\langle \omega^2 \rangle/dt|}
   \eqno{(14)}
$$
(actually, only one of the (14) inequalities is sufficient for further consideration). For the nonlinear 
system the principal problem is {\it invariance} of the characteristic time-scales in (14) to the rescaling 
$\psi \rightarrow \lambda \psi$ (otherwise the inequalities (14) become meaningless). It can be readily shown 
that the cancellation of the nonlinear terms in the 
dynamical equations (8),(9) provides such invariance for the enstrophy based time-scales in Eq. (14). 

Since for the transitional turbulence $\Omega^{ac}$, $\Omega^{pc}$, and $\langle \omega^2 \rangle$ are still of the 
same order, then we obtain from Eq. (14)
$$
\left| \frac{d\langle \omega^2 \rangle}{dt}\right| \ll \left|\frac{d\Omega^{ac}}{dt} \right|,~~~
\left| \frac{d\langle \omega^2 \rangle}{dt}\right| \ll \left|\frac{d\Omega^{pc}}{dt} \right| \eqno{(15)}
$$

Then from $\Omega^{ac}= \langle \omega^2 \rangle - \Omega^{pc}$ and Eq. (15) we obtain
$$
\frac{d\Omega^{pc}}{dt}\simeq -\frac{d\Omega^{ac}}{dt} \eqno{(16)}
$$
and we can define exchange rate of the enstrophy, $\varepsilon_{\omega}$ in full 
analogy with Eq. (12)
$$
\varepsilon_{{\bf \omega}} =  \left|
\frac{d\Omega^{ac}}{dt} \right|\simeq  \left|
\frac{d\Omega^{pc}}{dt} \right| \eqno{(17)}
$$
i.e. the adiabatic invariance of the enstrophy substitutes its exact invariance in the case of the transitional 
turbulence.

For the weak nonlinearity one can expect that the angle-averaged spectrum $N_k$ 
for sufficiently large $k$ is proportional to $\varepsilon_n$ (cf \cite{ldn},\cite{jsp}). 
This can be also valid for the transitional turbulence 
for sufficiently large $k$. Then with the additional dimensional parameter $ \varepsilon_{{\bf \omega}}$ (which has dimension $T^{-2}$) one obtains scaling law for sufficiently large $k$ from dimensional considerations
$$
N_k \sim \varepsilon_n \varepsilon_{{\bf \omega}}^{-1/2}~ k^{-1}   \eqno{(18)}
$$
The enstrophy exchange controls the particle exchange in this case. 

External dissipation or forcing can make the estimates (14) invalid and, hence, destroy this scenario (cf \cite{df},\cite{no}). 
Though, in the case of a sufficiently weak and linear external dissipation (forcing) the above consideration can 
be still valid. In this case the exact invariance of the total number of particles $N$ can be replaced by its 
adiabatic invariance (similarly to the enstrophy) and the inequalities similar to the Eq.(14) can be made. Due to 
linearity of the dissipation (forcing) these inequalities will be still invariant to the rescaling: 
$\psi \rightarrow \lambda \psi$.   

There exists another way to tackle the problem. The wave function can be considered as a vector on the complex plane. 
The density: $|\psi|^2$, gives magnitude of the vector, whereas the phase gives its direction. Conditional 
functional average over the directions of the vector can reduce the dynamics of the $|\psi|^2$ to 
a passive scalar one (cf \cite{bsr}). Then, spectrum like Eq. (18) can be obtained in 
analogy with classic fluid turbulence controlled by the enstrophy 
adiabatic invariance (cf \cite{km}).

\section{Comparison with numerical simulations}

In paper \cite{bs} a large scale {\it three}-dimensional numerical simulations of the Gross-Pitaevskii equation: 
with $\eta ({\bf x}) =0$ and $U_{eff}(|\psi|^2) \propto |\psi|^2$,  
was performed in order to reveal all three stages of the evolution from weak turbulence
to superfluid turbulence (the Bose-Einstein condensate) with a tangle of quantized vortices. 
While the observed evolution for $t < 600$ exhibits the well defined self-similar weak turbulence, 
the period $600 < t < 1000 $ was identified in \cite{bs} with the transitional turbulence. In Figs. 1 and 2 
one can see the cumulative number of particles $ \bar{N}_k  = \Sigma_{k'<k}~ N_k'$, 
which shows how many particles have momenta not exceeding $k$ at $t=600$ 
(the beginning of the transitional stage, Fig. 1) and at $t=1000$ (the end of the transitional 
stage, Fig. 2). 

Using scaling (18) one can estimate the number for sufficiently large $k$ as
$$
\bar{N}_k  = \Sigma_{k'<k}~ N_{k'} \sim  \Sigma_{k_c <k'<k}~ (k')^{-1} \sim \ln (k/k_c)  \eqno{(19)}
$$
where $k_c$ is the wavenumber scale separating (approximately) the pre- and above-condensate fractions.
We use the semi-log scales and the straight lines  in the Figs. 1,2 in order to indicate agreement 
with the Eq. (19). After the formation of the quasi-condensate at 
$t > 1000$ (third stage), the distribution of particles acquires a bimodal shape \cite{bs}. The value $k_c$ 
can be defined at crossing point of the straight line in Figs. 1,2 (indicating Eq. (19)) and 
the horizontal axis.  
\begin{figure} \vspace{-1.5cm}\centering
\epsfig{width=.45\textwidth,file=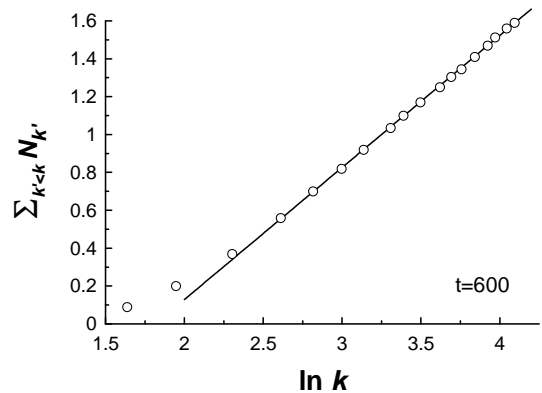} \vspace{-4.5cm}
\caption{The cumulative number of particles $\bar{N}_k  = \Sigma_{k'<k}~ N_{k'}$ against $\ln k$. The data 
are taken from \cite{bs} for $t=600$ (the beginning of the transitional stage). The straight line is drawn in 
order to indicate agreement with Eq. (19). }
\end{figure}
\begin{figure} \vspace{-0.5cm}\centering
\epsfig{width=.45\textwidth,file=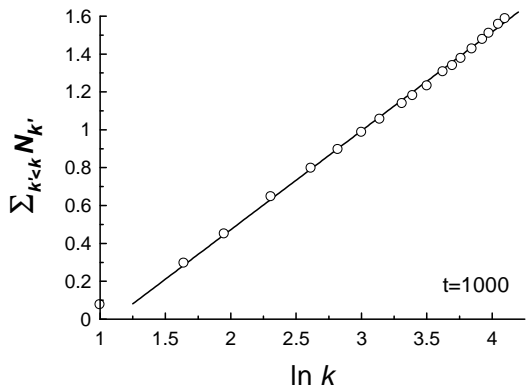} \vspace{-4.5cm}
\caption{As in Fig. 1 but for $t=1000$ (the end of the transitional stage).}
\end{figure}
\section{Visibility function and CMB}

So-called visibility function is used to describe 'optical' properties of the plasma at decoupling stage. 
The visibility function $g(t)$ is defined so that the probability that a CMB photon last scattered between time 
$t$ and $t+dt$ is given by $P(t)dt$. A measure of the width of the visibility function around its (rather sharp) 
maximum can be used for quantitative characterization of the decoupling stage. This stage is also called as 
last scattering shell (or for the CMB measurement purposes as last scattering surface). The last scattering 
probability turns out to be a narrow peak around a decoupling redshift. 
The visibility function is defined as 
$$
g(t) = n_e \sigma_T a(t) \exp  \left\{-\int_t^{t_0} n_e(t')\sigma_T a(t')dt'\right\}  \eqno{(20})
$$
where 
$a(t)$ is the expansion factor normalized to unity today, $n_e$ is the electron density, 
and $\sigma_T$ is the Thomson cross section. 
\begin{figure} \vspace{-0.5cm}\centering
\epsfig{width=.45\textwidth,file=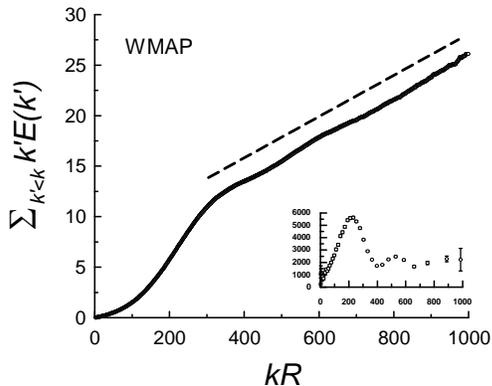} \vspace{-4.5cm}
\caption{The cumulative compensated spectrum of the CMB temperature fluctuations 
calculated using the WMAP 3-year data \cite{wmap}. The dash straight line in this figure 
indicates agreement with Eq. (27). The cumulative spectrum is given in arbitrary units. 
The insert shows the WMAP three-year {\it binned} spectrum $l(l+1)/2\pi C_l$ against $l$  
\cite{wmap} (see Eq. (32)). }
\end{figure}
The motion of the scatters imprints a temperature fluctuation,
$\delta T$, on the CMB through the Doppler effect 
$$
\frac{\delta T ({\bf n})}{T} \sim \int g(L){\bf n} \cdot {\bf v}
({\bf x})~dL  \eqno{(21})
$$
where ${\bf n}$ is the direction (the unit vector) on the sky and
${\bf v}$ is the velocity field of the electrons evaluated along
the line of sight, ${\bf x} = L{\bf n}$. Using Eq. (20) let us introduce 
a modified visibility function: $g'=g/n_e$. 
Then, 
$$
\frac{\delta T ({\bf n})}{T} \sim \int g'(L){\bf n} \cdot {\bf j}
({\bf x})~dL   \eqno{(22})
$$
where we used $n_e=|\psi|^2$ and ${\bf j}=|\psi|^2 {\bf v}$ is the weighted velocity filed. 
Unlike the potential velocity field ${\bf v}$ (Eq. (6)) the weighted velocity ${\bf j}$ has a 
rotational component in addition to the potential one ($\nabla \times {\bf j} \neq 0$, Eq. (7)). It is significant 
in present content due to the well known 'geometrical' cancellation of the contribution of the potential 
component into the integral (22) (see \cite{hd},\cite{vish}). The well known Helmholtz theorem decomposes 
any sufficiently smooth, decaying vector field into potential (curl-free) ${\bf P}$ and solenoidal (divergence-free) ${\bf S}$ component vector fields: ${\bf j} = {\bf P} + {\bf S}$, where
$$
{\bf S (r)} = \frac{1}{4\pi} \nabla \times \int \frac{\nabla' \times {\bf j}}{|{\bf r} - 
{\bf r'}|} d{\bf r'} \eqno{(23)}
$$  
so that
$$
\frac{\delta T ({\bf n})}{T} \sim \int g'(L){\bf n} \cdot {\bf S}
({\bf x})~dL   \eqno{(24})
$$
Since $\nabla \times {\bf j} = \omega |\psi|^2$, then taking time derivative of the both sides 
Eq. (23) and using Eq. (9) 
one can immediately conclude that dynamical equation for ${\bf S}$ does not contain the {\it nonlinear} terms 
explicitly (similarly to the equations (8) and (9)). Therefore, one can define an exchange rate of ${\bf S}^2$: 
$~\varepsilon_{{\bf s}}$, in full analogy with Eq. (17). If the weak nonlinear approximation is valid for 
the decoupling stage, then, the angle-averaged spectrum of the ${\bf S}$-field fluctuations 
$\langle {\bf S}^2 \rangle = \int \Phi_k dk$:
$$
\Phi_k \sim \varepsilon_{{\bf s}} \varepsilon_{{\bf \omega}}^{-1/2}~ k^{-1}   \eqno{(25)}
$$
can be analogously obtained for sufficiently large $k$. That means that the scaling $k^{-1}$ 
can be also observed for corresponding angle-averaged CMB temperature fluctuations spectrum $E(k)$
$$
\frac{\langle \delta T^2 \rangle}{T^2} = \int E(k) dk , ~~~ E(k) \sim k^{-1}  \eqno{(26)}
$$
for sufficiently large $k$. It follows from Eq. (26) that the cumulative {\it compensated} spectrum
$$
\sum_{k'<k} k'E(k') \simeq \sum_{k'<k_c} k'E(k') + \int_{k_c}^{k} k'E(k') dk' \simeq A+Bk \eqno{(27)}
$$
where $A$ and $B$ are certain constants. The cumulative spectrum calculated using the WMAP 3-year 
combined data (version 2.0, March 2006) is shown in figure 3. The dash straight line in this figure 
indicates agreement with Eq. (27) (the wavenumber $k$ is normalized by the comoving angular-diameter 
distance to the last scattering surface: $R$).

The resonant acoustic waves (see Introduction) present a finite-size effect for this scaling (actually, 
the so-called first acoustic peak represents the pre-condensate, see next Section). 
Therefore, these waves can be taken 
into account in a standard for the finite-size effects way \cite{sorn}:
$$
E(k) \sim f(k/k_c) k^{-1}  \eqno{(28)}
$$
where the function $f(k/k_c)$ of the dimensionless wavenumber $k/k_c$ corresponds to the finite-size effect. 
Then the properly compensated spectrum
$$
kE(k) \sim f(k/k_c)  \eqno {(29)}
$$
represents the acoustic-resonant effect after extraction of the scaling component (26).

In the CMB literature it is used to expand the CMB temperature
fluctuations in spherical harmonics
$$
\frac{ \delta T }{T} = \sum_{lm} a_{lm} Y_{lm} (\Theta, \phi)  \eqno{(30)}
$$
For isotropic situation the angular power spectrum of the fluctuations, $C_{l}$, is defined as 
$\langle a_{lm}a_{l'm'}^{*} \rangle = C_{l} \delta_{ll'}\delta{mm'}$. Then, for sufficiently large 
$l$
$$
\frac{\langle \delta T^2 \rangle}{T^2}= \frac{1}{4\pi} \sum_{l} (2l+1) C_l \simeq \frac{1}{2\pi} 
\int (l+1) C_l~ dl \eqno{(31)}
$$
(see for more details \cite{su}). It follows from Eqs. (26) and (31) that for sufficiently large $l$: 
$E(k) \propto (l+1)/2\pi~ C_l$, where the comoving wavenumber $k \simeq l/R$.
Hence, for sufficiently large $l$ the finite-size (i.e. acoustic-resonant) 
corrected scaling $E(k) \sim f(k/k_c) k^{-1}$ 
can be written in the compensated form $kE(k) \sim f(k/k_c)$: 
$$
l (l+1) C_{l} \simeq f(l/l_c)  \eqno{(32)}
$$
where $f(l/l_c)$ represents the acoustic-resonant peaks. It should be noted that in the 
{\it observed} angular power spectrum $C_{l}$ and even in the $(l+1) C_l$ (which represents spectrum $E(k)$ 
for sufficiently large $l \simeq kR$) one cannot see a hint on the first acoustic-resonant {\it peak} 
(see next Section). 
The only properly compensated spectrum $kE(k)$ (Eq. (29)) or $l (l+1) C_{l}$ (Eq. (32)) gives clear exhibition of 
the acoustic-resonant peaks (see the insert in Fig. 3). 

It is interesting to note that for small $l$ ($3<l<30$) the scale-invariant {\it potential} perturbations
generate CMB temperature fluctuations, which also have a nearly constant $l(l + 1)C_l$ 
(the so-called Sachs-Wolfe effect \cite{sw}). This makes the compensated spectrum $l(l + 1)C_l$ an adequate 
tool for a wide range of scales.

\section{Pre-condensate}

For the quantum Fermi plasmas an additional (nonlocal) nonlinearity was also considered in the form of a potential 
given by an additional Poisson equation \cite{m},\cite{se},\cite{ss}:
$$
{\rm i}\hbar \pd \psi t = -\left[ {\frac{\alpha\hbar^2}{2m}} \nabla^2 
 +\eta ({\bf x})+ \Phi({\bf x}, |\psi|^2)  
- U_{eff}(|\psi|^2) \right] \psi,   
$$
$$
\Delta \Phi = \frac{e}{\varepsilon} (|\psi|^2-n_0)  \eqno{(33)}
$$
where the potential $\Phi({\bf x}, |\psi|^2)$ is a functional of the density $|\psi ({\bf x})|^2$, 
$e$ is the electron charge, $\varepsilon$ the dielectric constant, and $n_0$ is the equilibrium electron number 
density. Splitting the potential $\Phi =\Phi_1+\Phi_2$, where 
$$
\Delta \Phi_1 = \frac{e}{\varepsilon} |\psi|^2,~~~ \Delta \Phi_2 = -\frac{e}{\varepsilon} n_0, \eqno{(34)}
$$ 
and using the integral representation of the potential $\Phi_1$:
$$
\Phi_1 ({\bf x}, |\psi|^2)= -\frac{e}{4\pi\varepsilon} \int \frac{|\psi({\bf x'})|^2}{|{\bf x} - {\bf x'}|} d{\bf x'} 
\eqno{(35)}
$$
it can be readily shown that the nonlinear Schr\"{o}dinger-Poisson system possesses the same type of 
mesoscopic gauge invariance as the Schr\"{o}dinger equation (1) discussed in the Section I. I.e. 
similarly to the nonlinear potential $U_{eff}(|\psi|^2)$ the nonlinear nonlocal potential $\Phi_1 ({\bf x}, |\psi|^2)$ does not appear explicitly in the 'classic' limit Eq. (2) (i.e. in the Ehrenfest theorem). Indeed,
$$
\langle \nabla \Phi_1 \rangle \propto \int \int  |\psi ({\bf x})|^2|\psi ({\bf x'})|^2 \nabla \left[ \frac{1}{|{\bf x}-{\bf x'}|}\right] d{\bf x}d{\bf x'}=
$$
$$
=\int \int  |\psi ({\bf x})|^2|\psi ({\bf x'})|^2~ \frac{({\bf x'}-{\bf x})}{|{\bf x}-{\bf x'}|^3} d{\bf x}d{\bf x'}=0 
  \eqno{(36)}
$$ 
\begin{figure} \vspace{-0.5cm}\centering
\epsfig{width=.45\textwidth,file=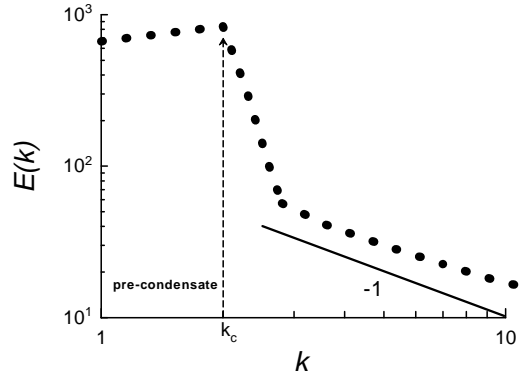} \vspace{-5.5cm}
\caption{Angular-average energy spectrum for a two-dimensional dense plasma. The data are taken 
from a numerical simulation presented in \cite{ss} for the Eq. (33) with weak nonlinear conditions. }
\end{figure}
\begin{figure} \vspace{-1.5cm}\centering
\epsfig{width=.45\textwidth,file=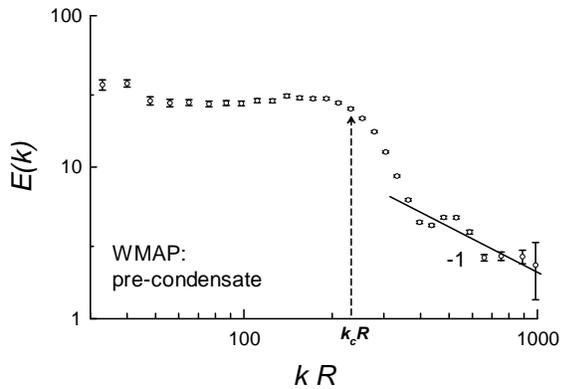} \vspace{-4.5cm}
\caption{Angular-average spectrum $E(k)$ (Eq. (37)) in the
wavenumber space for the 
binned WMAP data \cite{wmap} for $l = kR > 30$ (cf. the insert in Fig. 3). }
\end{figure}
In recent paper \cite{ss} results of a two-dimensional numerical simulation with a dense Fermi plasma 
have been reported. The authors of \cite{ss} have used the equations (33) with $U_{eff} (|\psi|^2) \propto 
|\psi|^2$. An energy spectrum obtained in this 
numerical simulation under weak nonlinear conditions is shown in Fig. 4 in the log-log scales. 
The authors of the Ref. \cite{ss} 
related the strong hump in the infrared part of the spectrum to a condensation process, which was also 
directly observed in this simulation. Although the causes for condensation and the coherent plasma oscillations 
in the dense Fermi plasma considered in \cite{ss} are rather different from those at the decoupling, 
the equations and the main properties of the condensation process are similar. Therefore, it is interesting to compare the results of the 
simulation with the CMB data. To make this comparison possible one needs in the spectrum $E(k)$ for the 
temperature fluctuations. As it is shown in the Section V for sufficiently large $l$: 
$$
E(k) \propto (kR+1)C_{kR}  \eqno{(37)}  
$$
where in the $(l+1)C_l$ the variable $l$ is taken as $l\approx kR$. Considering $l=30$ as a 
sufficiently large value for this purpose we show in Fig. 5 spectrum $E(k)$ (in the log-log scales) calculated 
using the binned WMAP data  \cite{wmap} (these data are shown 
in another presentation in the insert in Fig. 3). The arrow in Fig. 5 indicates the absolute maximum of the 
graph shown in the insert in Fig. 3
(the top of the first acoustic peak). Similarity between figures 4 and 5 is quiet obvious. It seems plausible to interpret the scale $k_cR$ (Fig. 5) as the 
scale separating between the pre- and above-condensate fractions at the decoupling (see Sections III,IV). This is also 
the {\it input}-scale of the external coherence at the decoupling (see Section I).  
The flatness of the pre-condensate fraction of the spectrum $E(k)$ (Fig. 5) indicates a rather advance stage of the 
condensation at the end of the decoupling process (a final part of the transition to the condensate state). 

\section{Discussion}

Different sources of turbulence in the early universe in the times preceding the decoupling 
were discussed in the literature, from the very early times \cite{g} up to the recombination time 
\cite{ber2}-\cite{beo}. In present paper we have discussed a possibility that the decoupling 
process itself can generate a turbulence related to a condensation process in the primordial plasmas. 
Just this 'later' turbulence has a preferable chance to imprint itself both on the CMB and 
on the further dynamics of the luminous matter. Especially significant is the question: whether the 
condensate itself with its tangles of quantized vortices appeared at the end of decoupling or the 
decoupling was finished at the transitional stage of the condensation process 
(see the end of the Section I) . The apparent 
flatness of the pre-condensate fraction of spectrum $E(k)$ in Fig. 5 (in the range: $k_cR > kR > 30$) 
indicates a rather advance stage of the condensation process in this case. One can even speculate that 
the end of the transition from the pre-condensate to the condensate is the effective end of the decoupling 
(and just this is imprinted in the CMB spectrum shown in Fig. 5). 
But the question is too serious to make definite conclusion from this observation only.

On the other hand, the recombination and dissipation 
processes could effect the decoupling turbulence. However, the coherent nature of this comparatively 
large-scale turbulence makes it stable to these effects. First of all, the recombination obviously is 
much slower process then the particles (enstrophy) exchange between the pre- and above-condensate 
fractions. Therefore, the total number of the particles $N$ can be considered as an adiabatic invariant 
at the decoupling and the consideration of the section III can be preserved when one takes into account the recombination. The linear dissipation can be treated 
by the same way (considering $N$ as an adiabatic invariant). \\

The author acknowledges the use of the Legacy Archive for
Microwave Background Data Analysis (LAMBDA) \cite{wmap}.

\end{document}